\documentclass[aps,prd,twocolumn,groupedaddress]{revtex4}
\usepackage{amsmath}

\begin{document}


\title{A hyperbolic tetrad formulation of the Einstein equations for numerical relativity}


\author{L. T. Buchman}
\affiliation{Astronomy Department, University of Washington, Seattle, WA}
\author{J. M. Bardeen}
\affiliation{Physics Department, University of Washington, Seattle, WA}


\date{\today}

\begin{abstract}
The tetrad-based equations for vacuum gravity published by
Estabrook, Robinson, and Wahlquist are simplified and
adapted for numerical relativity.  We show that the evolution equations
as partial differential equations for the Ricci rotation coefficients constitute
a rather simple first-order symmetrizable hyperbolic system, not only
for the Nester gauge condition on the acceleration and angular velocity of the
tetrad frames considered by Estabrook {\it et al.}, but also for the Lorentz
gauge condition of van Putten and Eardley and for a fixed
gauge condition.  We introduce a lapse function and a shift vector to allow general
coordinate evolution relative to the timelike congruence defined by the tetrad vector
field.
\end{abstract}


\maketitle

\setlength{\unitlength}{.15cm}
\section{Introduction}
Gravitational wave detection is soon to become a reality.  However, the scientific
community has not yet been able to calculate gravitational waveforms 
from the most likely source, namely the violent and dynamic 
merger phase of binary black hole collisions \cite{LL01}.  We hope to 
contribute to this effort
with detailed small-scale studies.  Our first such study of 1D colliding gravitational 
plane waves \cite{BaBu02} indicated that two important factors which improve 
the accuracy and stability of the numerical calculations are hyperbolicity 
of the equations and evolving variables which are related to physical quantities.  
As a way to generalize our results to black hole spacetimes and higher dimensions, 
we are investigating a tetrad approach based on the formalism published by 
Estabrook, Robinson, and Wahlquist (ERW) \cite{ERW97}.  
In this paper, we present a modified version of the formalism, adapted
for numerical relativity.  Subsequent papers will present our numerical results.

The standard approach to vacuum numerical relativity
is a $3+1$ decomposition, of the type introduced by Arnowitt-Deser-Misner \cite{ADM62}. 
The $3+1$ formulations slice four-dimensional spacetime into three-dimensional
spacelike hypersurfaces (see \cite{SY02} for a recent review).  They evolve the spatial metric and extrinsic curvature of the hypersurfaces,
which are expressed in a coordinate basis.  The evolution of the coordinates is
described by a lapse function and shift vector, which may or may not be dynamic.

A tetrad formulation uses orthonormal basis vectors, $\boldsymbol{e}_{\alpha}$ ($\alpha=0,\:1,\:2,\:3$), 
which describe local Lorentz frames.  The spacetime metric, the dot product of the basis vectors,
is everywhere the Minkowski metric, $g_{\alpha \beta}=\boldsymbol{e}_{\alpha }
\cdot\:\boldsymbol{e}_{\beta}=\eta_{\alpha \beta}$.  The timelike vector field of the
orthonormal frames, $\boldsymbol{e}_{0}$, defines a preferred
timelike congruence, to which it is tangent.  The spatial triad vectors
in a particular rotational orientation with respect to $\boldsymbol{e}_{0}$ are
$\boldsymbol{e}_{1},\:\boldsymbol{e}_{2}$, and $\boldsymbol{e}_{3}$.  The dual
basis of orthonormal one-forms is $\boldsymbol{e}^{\alpha}$ such that $\langle
\boldsymbol{e}^{\alpha},\:\boldsymbol{e}_{\beta}\rangle
=\delta^{\alpha}_{\beta}$ and $\boldsymbol{e}^{\alpha}
\cdot\:\boldsymbol{e}^{\beta}=\eta^{\alpha \beta}$.  As Estabrook 
and Wahlquist point out \cite{EW64}, these tetrad frames are natural for 
measuring observable physical quantities.

Most of the recent tetrad formalisms assume that the 
tetrads are tied to the physically defined flow of a fluid,
primarily in the context of cosmology or of interior metrics of rotating
stars.  Here, we advocate that a tetrad formalism may be useful even in vacuum blackhole 
spacetimes.  The variables in a tetrad formalism are the connection coefficients (often called the
Ricci rotation coefficients), the tetrad vector components, and, typically, the
tetrad components of the Weyl tensor or the Riemann tensor.  The spatial coordinates are often assumed from
the start to be comoving with the timelike congruence generated by the tetrad field. 
In dealing with black hole event horizons, however, it is important
to allow general choices of coordinates.  We do this by introducing a lapse function
and a shift vector defined relative to the congruence world lines (see also van Putten
and Eardley \cite{VPE96}).

The ERW formulation is a quasi-FOSH (first order symmetric hyperbolic) system.
We use the term ``quasi'' because the directional derivatives along the spatial tetrad legs 
contain partial time derivatives (also see \cite{CBY02}).  The basic quasi-FOSH structure of the
equations involves the Weyl tensor components as variables.  The quasi-FOSH system also includes
equations derived from the Nester gauge conditions \cite{JN92}.  These equations evolve the acceleration
of the congruence worldlines, $\boldsymbol{a}$, and the angular velocity of the spatial tetrad
vectors relative to Fermi-Walker transport, $\boldsymbol{\omega}$.  The tetrad components
$a_a$ and $\omega_a$ are gauge quantities since the spacetime orientation of the tetrad is not
fixed by the spacetime geometry.  Both $\boldsymbol{a}$ and $\boldsymbol{\omega}$ are orthogonal to
$\boldsymbol{e}_{0}$, and therefore have only the three spatial tetrad components.
In addition to evolution equations, the Nester gauge conditions provide equations
which constrain the spatial dependence of the $a_a$ and the $\omega_a$ at
any given time.  

ERW's formalism can be modified to give
a particularly simple quasi-FOSH system, as suggested by Estabrook and Wahlquist (EW) \cite{EW02},
by eliminating the Weyl tensor components as separate variables,
and adding the Nester constraint equations to the evolution equations.

In this paper, we discuss the derivation of the EW formalism in a way which is perhaps
more accessible to those familiar with standard tensor analysis, as opposed to
Cartan differential form analysis.  What distinguishes the EW formalism is that the basic
quasi-FOSH structure involves only the connection coefficients.  Most of the tetrad formulations
in the literature include the Weyl tensor or the Riemann tensor to get a simple quasi-hyperbolic structure. 
Unless one adds the constraints to the Einstein equations in the particular way we present in this paper,
the quasi-hyperbolic system involving only the connenction coefficients is quite complicated.  We do not
claim that the EW system is simpler or more elegant that those involving the Weyl or Riemann tensor; however,
it has fewer variables and fewer constraint modes, which may be advantageous in 3D codes.

After deriving the EW formalism, we proceed to extend it
into a form useful for vacuum numerical relativity.  We consider two gauge conditions
other than the Nester gauge which also give 
quasi-FOSH systems of equations.  These are
the Lorentz gauge used by van Putten and Eardley
\cite{VPE96}, and a fixed gauge, where $a_a$ and $\omega_a$ 
are fixed functions of time and space.  Additionally, we allow
for a completely arbitrary relationship between the congruence
and the hypersurface.  Finally, we analyze the true hyperbolicity
of the equations when expressed in terms of partial derivatives.
The partial differential system contains evolution equations for
eighteen non-gauge connection coefficients,
six gauge quantities (if a dynamic gauge condition
is chosen), three components of a vector describing the velocity of the congruence
relative to the hypersurface, and nine components of the spatial tetrad vectors projected into
a $t=$ constant hypersurface.  These last twelve variables, together with our
lapse function and shift vector (which we do not evolve), completely determine
the sixteen tetrad vector components.  The flux Jacobian of the partial differential 
system of equations has a complete set of eigenvectors and
real eigenvalues, thus satisfying the requirement for hyperbolicity as per LeVeque \cite{LeV02}.
Furthermore, these equations are symmetrizable hyperbolic as defined by 
Lindblom and Scheel \cite{LS02}. 

Tetrad formulations for general relativity other than ERW and EW
include those by Friedrich \cite{HF96}, Jantzen, Carini, and Bini \cite{Jantzen01}, 
van Elst and Uggla \cite{VEU97},
Choquet-Bruhat and York \cite{CBY02}, and van Putten and Eardley \cite{VPE96}. 
Except for EW, all of these systems include the Weyl or Riemann tensor components as fundamental variables.  
Friedrich's paper is a definitive
discussion of hyperbolicity for both tetrad and $3+1$ representations.
Janzten, Carini, and Bini give an extensive historical review of the tetrad and
$3+1$ approaches.  They provide a unified framework for 
the two approaches in what they call ``gravitoelectromagnetism'', in which, however,
they only consider co-moving spatial coordinates.  van Elst and Uggla's formalism 
applies only to non-rotating congruences and orthogonal
hypersurface slicings.  Also, they do not bring the equations into a
quasi-FOSH form.  The main emphasis of Choquet-Bruhat and York's paper is a tetrad formulation
for fluids, where the congruence is aligned with the fluid
flow lines, and the spatial coordinates are co-moving
with the congruence.  The acceleration of the congruence worldlines is given
by the acceleration of the fluid,
and the angular velocity variables (our $\omega_a$'s) are fixed functions.  
Choquet-Bruhat and York do briefly mention the vacuum case,
in which their system is quasi-FOSH for arbitrary given acceleration and angular
velocity.  van Putten and Eardley use the Lorentz gauge condition
to obtain second-order wave equations for the connection coefficients.
van Putten gives a first-order form of these equations, involving the Riemann
tensor as well as the connection coefficients, for his numerical implementation in 1D 
Gowdy wave vacuum spacetime \cite{VP97}.  The van Putten-Eardley 
formulation allows for complete freedom in the choice of spacetime foliation.

An example of using orthonormal frames in spatial hypersurfaces is the Ashtekar formulation
(with subsequent modifications) \cite{AA86, AA87, ILR98, YS99, YS00}.
Ashtekar follows a $3+1$ split in the sense that his variables are defined
relative to a hypersurface.  However, his description of the geometry of the
hypersurface is in terms of orthonormal triads instead of the metric.
He uses complex variables to give a compact formalism,
requiring the use of reality constraints to recover real spacetime.
As subsequently modified by Yoneda and Shinkai, Ashtekar's formulation
becomes a FOSH system of PDE's.  Shinkai and Yoneda \cite{SY00, YS01} have 
published numerical studies using this formulation in 1D planewave vacuum spacetime.

\section{\label{variables}Variables}

Throughout this paper, lower case
Greek letters denote spacetime indices (0-3), and lower case Latin letters denote only spatial indices
(1-3).  The letters in the beginning of the alphabets, ($\alpha,\:\beta,\:\gamma,\:\delta,\:\epsilon$) and 
(a, b, c, d, e, f), denote tetrad indices.  Mid-alphabet letters 
($\lambda,\:\mu,\:\nu$) and (i, j, k, l) denote coordinate indices.  
Repeated indices are summed in all cases.

For an orthonormal tetrad of basis vector fields, there are twenty-four distinct 
connection coefficients.  These coefficients are scalar fields under coordinate
transformations, and are defined as 
\begin{equation}
\label{defngamma}
\Gamma_{\alpha\beta\gamma}=\boldsymbol{e}_{\alpha}\cdot\nabla_{\gamma}\:\boldsymbol{e}_{\beta}=-\Gamma_{\beta \alpha \gamma},
\end{equation}
with $\nabla$ the covariant derivative operator.  The $\Gamma_{\alpha\beta\gamma}$
are the same as Ricci rotation coefficients (see Wald \cite{Wald84}).  They can
be written in terms of commutators of the basis vectors:
\begin{multline}
\label{conncomm}
\Gamma_{\alpha\beta\gamma}=\frac{1}{2}\{\langle\boldsymbol{e}^{\delta},\:[\boldsymbol{e}_{\alpha},\:\boldsymbol{e}_{\beta}]\rangle\:\eta_{\delta\gamma}+\\
\langle\boldsymbol{e}^{\delta},\:[\boldsymbol{e}_{\alpha},\:\boldsymbol{e}_{\gamma}]\rangle\:\eta_{\delta\beta}-\langle\boldsymbol{e}^{\delta},\:[\boldsymbol{e}_{\beta},\:\boldsymbol{e}_{\gamma}]\rangle\:\eta_{\delta\alpha}\}.
\end{multline}

Just as in the $3+1$ formalisms, it is convenient to make a space-time split in the
tetrad formulation.  Here, however, the split is relative to the timelike congruence 
defined by the tetrad rather
than the constant-t spacelike hypersurface.  The connection coefficients can then be 
re-labeled as 3D quantities with spatial triad indices (see Wahlquist \cite{HW92} and ERW).  To begin,
\begin{equation}
K_{ba}\equiv\Gamma_{a0b}, 
\end{equation}
where the symmetric part of $K_{ba}$ is the rate of strain of the congruence, and the antisymmetric
part,
\begin{equation}
\Omega_{a}\equiv\frac{1}{2}\:\varepsilon_{abc}\:K_{bc},
\end{equation}
is the vorticity of the congruence.
If $\Omega_a=0$, the congruence is hypersurface orthogonal, and 
$K_{ba}$ is the traditional extrinsic curvature of the orthogonal hypersurface. 
Note that the sign of $K_{ba}$ here is the same as in ERW and Wald, but opposite to
that of Misner, Thorne, and Wheeler \cite{MTW73}. 

The spatial tetrad connection coefficients can be expressed more compactly as a two-index
quantity defined by
\begin{equation}
N_{ba}\equiv\frac{1}{2}\:\varepsilon_{acd}\:\Gamma_{cdb}.
\end{equation}
The diagonal components, $N_{11}$, $N_{22}$, and $N_{33}$,
describe the twists of the spatial triads along the
$1$, $2$, or $3$ directions, respectively.  The combinations
$N_{ab}+N_{ba}$ of the non-diagonal components represent gravitational wave degrees of
freedom.  It is sometimes convenient to represent the antisymmetric part of $N_{ba}$ by its
own symbol
\begin{equation}
n_{a}\equiv\frac{1}{2}\:\varepsilon_{abc}\:N_{bc}.
\end{equation}

The acceleration of the congruence is
\begin{equation}
a_{a}\equiv\Gamma_{a00},
\end{equation}
and the angular velocity of the spacelike triads relative to Fermi propagated axes is
\begin{equation}
\omega_{a}\equiv\frac{1}{2}\:\varepsilon_{abc}\:\Gamma_{cb0}.
\end{equation}

There are nine $K_{ba}$ and nine $N_{ba}$, giving eighteen primary variables
to be evolved.  The three $a_{a}$ and three $\omega_{a}$ are gauge quantities, 
which, in this paper, are evolved by either the Nester or Lorentz gauge, or kept fixed.

Numerical calculations are performed with a particular choice of coordinates $x^{\mu}$. 
Hyperbolic evolution consists in calculating variables on the spacelike hypersurfaces
characterized by $x^{0}=t_2$ from an initial state specified by values of the variables
on an earlier spacelike hypersurface $x^{0}=t_1$.  Let $\lambda_{\alpha}^{\mu}$
denote the transformation matrix between coordinate basis vectors and the tetrad
basis vectors:
\begin{equation}
\label{ealpha}
\boldsymbol{e}_{\alpha}=\lambda_{\alpha}^{\mu}\:\boldsymbol{e}_{\mu}.
\end{equation}
From Eq.~(\ref{ealpha}), we obtain directional derivatives along the tetrad 
directions in terms of the partial derivatives along the coordinate directions,
\begin{equation}
\label{directional}
D_{\alpha}=\lambda_{\alpha}^{\mu}\:\frac{\partial}{\partial x^{\mu}}.
\end{equation}
The $\lambda_{\alpha}^{\mu}$ are the coordinate components of the tetrad basis vectors. 
The coordinate metric is constructed from these tetrad vector components as 
$g^{\mu\nu}=\eta^{\alpha\beta}\:\lambda_{\alpha}^{\mu}\:\lambda_{\beta}^{\nu}$.

We find it simpler to use as variables not the sixteen $\lambda_{\alpha}^{\mu}$,
but the following sixteen quantities which completely determine the tetrad vector components:
nine $B_a^k$, the coordinate components of
projections of the spatial tetrad vectors into the hypersurface; three
$A_a$, which measure the tilts of the spatial tetrad vectors relative to
the hypersurface and are (minus) the tetrad components of the 3-velocity
of the hypersurface frame relative to the tetrad frame; the tetrad lapse
function $\alpha$ and the three coordinate components of the tetrad shift
vector $\beta^k$, which describe the evolution of the coordinates relative
to the tetrad congruence.  In this paper we evolve the $B_a^k$ and the $A_a$
as dynamic variables, but take $\alpha$ and the $\beta^k$ to be fixed
functions of the coordinates.  Eventually we may want to expand the
hyperbolic system to include dynamic equations for the lapse and the
shift.

In a $3+1$ formalism, the lapse function $N$ is the rate of change of
proper time with respect to coordinate time along the hypersurface normal, and the
shift vector $N^k$ is the rate of displacement of the spatial coordinates
with respect to coordinate time relative to the hypersurface normal, such that the
coordinate velocity of the normal worldline $dx^k/dt = -N^k$.  Our tetrad lapse
function $\alpha$ is the rate of change of proper time with respect to coordinate time
along the tetrad congruence, and our tetrad shift vector $\beta^k$ is the rate of
displacement of the spatial coordinates relative to the tetrad congruence worldlines
per unit coordinate time.  For simple comoving coordinates, the
spatial coordinates are constant along the congruence, and $\beta^k = 0$.
However, comoving coordinates are not desirable in black hole
calculations, since with a finite acceleration tetrad worldlines will be
continuously advected inward across the event horizon.  The tetrad lapse
must be chosen so that the constant-time hypersurfaces remain spacelike,
which is equivalent to the condition $A_a\:A_a < 1$.
        
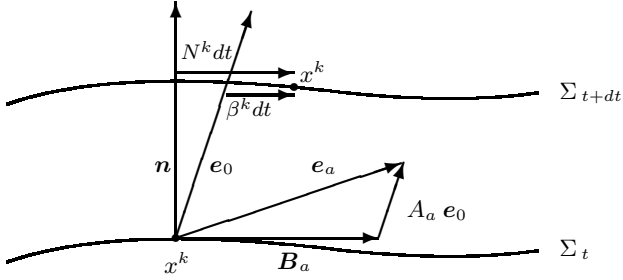
\begin{figure}[h]
\begin{picture}(50,33)\thicklines
\qbezier(0,4)(10,7.5)(30,5)
\qbezier(30,5)(40,4)(47,5)
\qbezier(0,18)(10,21.5)(30,19)
\qbezier(30,19)(40,18)(47,19)
\put(15,6.1){\vector(0,1){21}}
\put(15,6.1){\vector(1,3){6.7}}
\put(15,6.1){\vector(3,1){20}}
\put(49,4.5){$\Sigma_{\:t}$}
\put(49,18.5){$\Sigma_{\:t+dt}$}
\put(25.5,19.5){\circle*{0.6}}
\put(26,20){$x^k$}
\put(15,6.1){\circle*{0.6}}
\put(14,3){$x^k$}
\put(27,12){$\boldsymbol{e}_a$}
\put(33,6.2){\vector(1,3){2.2}}
\put(35.5,8){$A_a\:\boldsymbol{e}_0$}
\put(15,6.1){\vector(1,0){18}}
\put(24,3.4){$\boldsymbol{B}_a$}
\put(18,12){$\boldsymbol{e}_0$}
\put(13.1,12){$\boldsymbol{n}$}
\put(19.5,18.8){\vector(1,0){6}}
\put(19.5,16.9){\footnotesize $\beta^kdt$}
\put(15,20.8){\vector(1,0){10.5}}
\put(15.45,22){\footnotesize $N^kdt$}
\end{picture}
\caption{\label{fig1}Decomposition of $\boldsymbol{e}_a$ into a vector
tangent to the hypersurface, $\boldsymbol{B}_a$, and a vector parallel to the congruence, 
$A_a\:\boldsymbol{e}_0$.  Displacements of the spatial coordinates, $x^k$, relative
to the congruence worldline
equal $\beta^k dt$, where $\beta^k$ is our tetrad shift vector.  Displacements of $x^k$ relative
to the hypersurface normal $\boldsymbol{n}$ are $N^k dt$, where $N^k$ is the $3+1$ shift vector.}
\end{figure}

The projection of $\boldsymbol{e}_a$ into the hypersurface is done along the
congruence worldlines, such that
\begin{equation}
\label{ea}
\boldsymbol{e}_a = {A}_a\:\boldsymbol{e}_0+\boldsymbol{B}_a.
\end{equation}
Fig. \ref{fig1} shows how these various vectors are related from the point of
view of a frame at rest with respect to the hypersurface.

The directional derivative along a spatial tetrad direction is
\begin{equation}
\label{Da}
D_a = {A}_a\:D_0+B_a^k\:\frac{\partial}{\partial x^k},
\end{equation}
and the directional derivative along the congruence is
\begin{equation}
\label{D0}
D_0=\frac{1}{\alpha}\:\Big(\frac{\partial}{\partial t}-{\beta}^k\:\frac{\partial}{\partial x^k}\Big).
\end{equation}
Plugging Eq.~(\ref{D0}) into Eq.~(\ref{Da}), we get
\begin{equation}
\label{Daagain}
D_a=\Big(B_a^k-\frac{{A}_a}{\alpha}\:{\beta}^k\Big)\:\frac{\partial}{\partial x^k}+\frac{{A}_a}{\alpha}\:\frac{\partial}{\partial t}.
\end{equation}
Comparing Eqs.~(\ref{D0}) and (\ref{Daagain}) with Eq.~(\ref{directional}),
we can read off the tetrad vector components.  They are
\begin{equation}
\label{tetradmatrices}
\lambda_{\alpha}^{\mu}=
\begin{bmatrix}
\lambda_0^0 & \; & \lambda_0^k\\
\; & \;\\
\lambda_a^0 & \; & \lambda_a^k
\end{bmatrix}
=
\begin{bmatrix}
\frac{1}{\alpha} & \; & -\frac{{\beta}^k}{\alpha}\\
\; & \;\\
\frac{{A}_a}{\alpha} & \; & (B_a^k-\frac{{A}_a}{\alpha}\:{\beta}^k)
\end{bmatrix}.
\end{equation}

The $3+1$ lapse function, $N$, and shift vector components, $N^k$, can be expressed in terms of $\alpha$, ${\beta}^k$,
$A_a$, and $B_a^k$ by constructing $g^{tt}$ and $g^{tk}$ from the tetrad vector components using
$g^{\mu\nu}=\eta^{\alpha\beta}\:\lambda_{\alpha}^{\mu}\:\lambda_{\beta}^{\nu}$.  With the
relations
\begin{equation}
g^{tt}=-\frac{1}{N^2} \quad\mbox{and}\quad g^{tk}=-\frac{N^k}{N^2},
\end{equation} 
we get
\begin{equation}
N=\frac{\alpha}{\sqrt{1-A_a\:A_a}}\:,
\end{equation}
\begin{equation}
N^k=\:{\beta}^k\:+\:\frac{\alpha\:A_a\:B_a^k}{1-A_a\:A_a}\:.
\end{equation}
The tetrad lapse is smaller than the 3+1 lapse due to the time dilation
of the tetrad observer in the rest frame of the hypersurface.

\section{\label{reordering}The tetrad equations}

In this section, we present the evolution and constraint
equations for the connection coefficients defined in Sec. \ref{variables}
in terms of directional derivative operators along the
tetrad directions.  The structure of the equations in
this form is deceptively simple.  One has to keep in mind
that partial differential equations are solved in numerical
relativity, so the directional derivatives have to be
expanded into partial derivatives using Eqs.~(\ref{D0}) and (\ref{Daagain}). 
In terms of directional derivatives, the equations are quasi-FOSH.
We will present the necessary steps to obtain the true
hyperbolic form in Sec. \ref{hyperbolic}.  Furthermore, we call the ``constraint'' equations
presented in this section quasi-constraint equations, as is done in
\cite{CBY02}, because although they contain only spatial directional
derivatives, these spatial directionals contain time partials (see Eq.~(\ref{Daagain})). 
The details of converting the quasi-constraint equations into true constraint equations
are given in Appendix \ref{appA}.

In both the tetrad and $3+1$ approaches, there is an 
ordering ambiguity inherent in the derivation of a first order 
hyperbolic evolution system from the Einstein equations.
In first order $3+1$ formalisms, 
the spatial derivatives of the metric are independent variables, 
$D_{jkl}$ (which equal $\frac{1}{2} \partial_{j} h_{kl}$, where $h_{kl}$ is the spatial
metric).  Index re-ordering of the Riemann tensor is an interchange of spatial partial
derivatives such that 
\begin{equation}
\label{kst}
\partial_{i} D_{jkl}=\partial_{j} D_{ikl}. 
\end{equation}
Exploiting this freedom leads to a wide variety of
hyperbolic formulations \cite{KST}.  The standard
energy and momentum constraint equations can also be added to the evolution
system to get additional hyperbolic or non-hyperbolic structures.
In the context of a tetrad approach, the connection coefficients
are given in terms of the commutation relations, Eq.~(\ref{conncomm}). 
To derive integrability conditions from Eq.~(\ref{conncomm})
is messy.  We find it is much easier to approach the question
of ordering ambiguity from the symmetries of the Riemann tensor,
especially since these symmetries are explicit when the indices are all up or down,
and it is trivial to raise and lower indices with the Minkowski metric.
We do not consider a whole range of schemes as do Kidder, Scheel, and Teukolsky \cite{KST},
but rather a particular scheme which leads to a simpler hyperbolic structure than the others. 
To focus on the quasi-FOSH structure of the equations, we present only the principal terms here.
The second order source terms are given in Appendix \ref{appA}.

The Riemann tensor projected onto a tetrad is
\begin{eqnarray}
\label{riemanntensor}
R_{\alpha \beta \gamma \delta}=\boldsymbol{e}_{\alpha}\cdot(\nabla_{\gamma}\nabla_{\delta}-\nabla_{\delta}\nabla_{\gamma}-\nabla_{[\boldsymbol{e}_{\gamma},\boldsymbol{e}_{\delta}]})\:\boldsymbol{e}_{\beta} \nonumber \\
=D_{\gamma}\:\Gamma_{\alpha \beta \delta}-D_{\delta}\:\Gamma_{\alpha \beta \gamma}+\Gamma_{\alpha \epsilon \gamma}\:\Gamma^{\epsilon}_{~ \beta \delta }-  \nonumber \\*
\Gamma_{\alpha \epsilon \delta}\:\Gamma^{\epsilon}_{~ \beta \gamma}+\Gamma_{\alpha \beta \epsilon}\:(\Gamma^{\epsilon}_{~ \gamma \delta}-\Gamma^{\epsilon}_{~ \delta \gamma}), \nonumber \\
\end{eqnarray}
where $D_{\alpha}$ represents directional derivatives along the tetrad directions. 
The antisymmetry of $R_{\alpha \beta \gamma \delta}$ on the second pair
of indices is explicit in Eq.~(\ref{riemanntensor}).  The antisymmetry on the first pair is
also trivial because of the antisymmetry of $\Gamma_{\alpha \beta \gamma}$ on the first two indices. 
However, the Riemann identities
\begin{equation}
\label{symmetry}
R_{\alpha \beta \gamma \delta}=R_{\gamma \delta \alpha \beta},
\end{equation}
\begin{equation}
\label{cyclic}
R_{\alpha \beta \gamma \delta}+R_{\alpha \delta \beta \gamma}+R_{\alpha \gamma \delta \beta}=0
\end{equation}
lead to new Riemann constraints, which we exploit in the following.

First we derive all the possible quasi-constraint equations, 
noting the use of Eq. 14.7 from Misner, Thorne, and Wheeler.  The energy quasi-constraint equation is
\begin{equation}
\label{energyconstraint}
G_{00}=R_{1212}+R_{2323}+R_{3131}\sim 2\:D_{a}\:n_{a}.
\end{equation}
An analogous quasi-constraint equation 
for $\Omega_{a}$ is derived using the cyclic identity, Eq.~(\ref{cyclic}), on 
$0abc$:
\begin{equation}
0=R_{0213}+R_{0321}+R_{0132}\sim 2\:D_{a}\:\Omega_{a}.
\end{equation}
The momentum quasi-constraint equations are obtained from $G_{0a}=0$:
\begin{eqnarray}
\lefteqn{G_{01}=R_{0212}+R_{0313}}\nonumber\\
 & & \sim -D_{1}\:(K_{22}+K_{33})+D_{2}\:K_{12}+D_{3}\:K_{13},
\end{eqnarray}
\begin{eqnarray}
\lefteqn{G_{02}=R_{0121}+R_{0323}}\nonumber\\
 & & \sim -D_{2}\:(K_{11}+K_{33})+D_{1}\:K_{21}+D_{3}\:K_{23},
\end{eqnarray}
\begin{eqnarray}
\label{momen}
\lefteqn{G_{03}=R_{0131}+R_{0232}}\nonumber\\
  & & \sim -D_{3}\:(K_{11}+K_{22})+D_{1}\:K_{31}+D_{2}\:K_{32}. 
\end{eqnarray}
Similar quasi-constraint equations involving $N_{ab}$ 
are derived solely from Eq.~(\ref{symmetry}) applied to spatial
indices of the Riemann tensor:
\begin{equation}
R_{1213}-R_{1312}\sim D_{1}\:(N_{22}+N_{33})-D_{2}\:N_{12}-D_{3}\:N_{13},
\end{equation}
\begin{equation}
R_{2321}-R_{2123}\sim D_{2}\:(N_{11}+N_{33})-D_{1}\:N_{21}-D_{3}\:N_{23},
\end{equation}
\begin{equation}
R_{3132}-R_{3231}\sim D_{3}\:(N_{11}+N_{22})-D_{1}\:N_{31}-D_{2}\:N_{32}.
\end{equation}

We now to turn to evolution equations.  By using  Eq.~(\ref{symmetry}) in the momentum
quasi-constraint equations, we calculate {\it evolution} equations for the non-diagonal 
components of $N_{ab}$.  For example, interchanging the first and second
pairs of indices in Eq.~(\ref{momen}) as shown below gives evolution 
equations for $N_{21}$ and $N_{12}$:
\begin{eqnarray}
\lefteqn{G_{03}=R_{0131}+R_{3202}} \nonumber \\
  & & \sim -D_{0}\:N_{21}-D_{2}\:\omega_{1}-D_{3}\:K_{11}+D_{1}\:K_{31},
\\
\lefteqn{G_{03}=R_{3101}+R_{0232}} \nonumber \\
  & & \sim D_{0}\:N_{12}+D_{1}\:\omega_{2}+D_{2}\:K_{32}-D_{3}\:K_{22}.
\end{eqnarray}
Evolution equations for the diagonal components of $N_{ab}$ are obtained
solely from applications of Eq.~(\ref{symmetry}):
\begin{equation}
R_{2301}-R_{0123}\sim D_{0}\:N_{11}+D_{1}\:\omega_1+D_{2}\:K_{31}-D_{3}\:K_{21},
\end{equation}
\begin{equation}
R_{3102}-R_{0231}\sim D_{0}\:N_{22}+D_{2}\:\omega_2+D_{3}\:K_{12}-D_{1}\:K_{32},
\end{equation}
\begin{equation}
R_{1203}-R_{0312}\sim D_{0}\:N_{33}+D_{3}\:\omega_3+D_{1}\:K_{23}-D_{2}\:K_{13}.
\end{equation}

Evolution equations for $K_{ab}$ are obtained from the Einstein equations.
As is done elsewhere (see for example \cite{BM97}), a multiple
of the energy quasi-constraint equation times the spatial metric can be subtracted
from the $K_{ab}$ evolution equations.  In the formulation we are presenting,
this amounts to subtracting a multiple of Eq.~(\ref{energyconstraint}) from
the diagonal spatial components of the Ricci tensor,
since our metric is just $\delta_a^b$.  The same type of
procedure in $3+1$ formulations affects the evolution equations for the
non-diagonal as well as the diagonal components of $K_{ab}$, since the
spatial metric there is, in general, non-diagonal.
The number we choose to multiply the energy quasi-constraint equation by
is $1$, to bring the evolution equations for the diagonal $K_{ab}$ into
the same form as those for the non-diagonal components.  With the aid 
of Eq.~(\ref{symmetry}), we derive sample evolution equations for
$K_{ab}$ below:
\begin{eqnarray}
\lefteqn{R_{12}=-R_{1020}+R_{1323}=-R_{1020}+R_{2313}} \nonumber \\
  & & \sim D_{0}\:K_{21}-D_{2}\:a_{1}-D_{3}\:N_{11}+D_{1}\:N_{31},
\\
\lefteqn{R_{12}=-R_{2010}+R_{1323}} \nonumber \\
  & & \sim D_{0}\:K_{12}-D_{1}\:a_{2}-D_{2}\:N_{32}+D_{3}\:N_{22},
\end{eqnarray}
\begin{eqnarray}
\lefteqn{R_{11}-G_{00}=-R_{0101}-R_{2323}} \nonumber \\
  & & \sim D_{0}\:K_{11}-D_{1}\:a_{1}-D_{2}\:N_{31}+D_{3}\:N_{21}.
\end{eqnarray}

If the gauge quantities $a_a$ and $\omega_a$ are fixed functions
of time, then the quasi-FOSH structure is now complete and is represented
by the following system of eighteen equations for the nine $K_{ab}$ and nine
$N_{ab}$:
\begin{equation}
\label{kabevoln}
D_{0}\:K_{ab}-D_{a}\:a_{b}-\varepsilon_{acd}\:D_{c}\:N_{db}=S\line(1,0){1}K_{ab},
\end{equation}
\begin{equation}
\label{nabevoln}
D_{0}\:N_{ab}+D_{a}\:\omega_{b}+\varepsilon_{acd}\:D_{c}\:K_{db}=S\line(1,0){1}N_{ab},
\end{equation}
where $S\line(1,0){1}K_{ab}$ and $S\line(1,0){1}N_{ab}$ are source terms quadratic
in the connection coefficient variables.  (See Appendix \ref{appB}.)

Alternatively, one can implement a dynamic gauge.
Both the Nester gauge used by ERW and EW
and the Lorentz gauge used by van Putten and Eardley
result in evolution equations for 
$a_a$ and $\omega_a$ which , when added to Eqs.~(\ref{kabevoln}) and 
(\ref{nabevoln}), form a quasi-FOSH system.  

The Nester gauge conditions in \cite{JN92} are defined for an arbitrary number of dimensions.
In 4D spacetime, the Nester conditions state that two 1-forms, $\tilde{\boldsymbol q}$
and $\hat{\boldsymbol q}$, whose tetrad components are
\begin{equation}
\tilde{q}_\alpha=\:\Gamma_{\alpha\:\:\:\gamma}^{\:\:\:\gamma},\:\:\:\hat{q}_\alpha=
\:\varepsilon_{\alpha\beta\gamma\delta}\:\Gamma^{\beta\gamma\delta},
\end{equation}
are closed.  This implies vanishing exterior derivatives so that, in tetrad component form,
the Nester conditions are
\begin{equation}
\nabla_\alpha\:\tilde{q}_{\:\beta}
-\nabla_\beta\:\tilde{q}_{\:\alpha}=0,
\end{equation}
\begin{equation}
\nabla_\alpha\:\hat{q}_{\:\beta}
-\nabla_\beta\:\hat{q}_{\:\alpha}=0.
\end{equation}
The Nester conditions result in six evolution equations, 
\begin{eqnarray}
\label{nesterevoln}
D_{0}\:a_{b}-D_{c}\:K_{cb}=-(\omega_{c}-2\:\Omega_{c})\:N_{cb}+(Tr N)\:\omega_{b},\nonumber \\
D_{0}\:\omega_{b}+D_{c}\:N_{cb}=-(\omega_{c}-2\:\Omega_{c})\:K_{cb}-(Tr N)\:a_{b},
\end{eqnarray}
and six quasi-constraint equations,
\begin{align}
\label{nesterconstraint}
\varepsilon_{abc}\:D_{b}\:(a_{c}-2\:n_{c})&=2\:(Tr K)\:\Omega_{a}-(Tr N)\:(a_{a}-2\:n_{a})+ \nonumber\\
(a_{d}-2\:n_{d})\:N_{da},\nonumber \\
\varepsilon_{abc}\:D_{b}\:(\omega_{c}-2\:\Omega_{c})&=-(Tr N)\:\omega_{a}+(\omega_{d}-2\:\Omega_{d})\:N_{da}. 
\end{align}

The Lorentz gauge condition in \cite{VPE96} is $\nabla^{\gamma}\omega_{\gamma \alpha \beta}=0$, where
$\omega_{\gamma \alpha \beta}=\Gamma_{\alpha \beta \gamma}$.  For fixed $\alpha$ and $\beta$, 
the $\omega_{\gamma \alpha \beta}$
are the components of connection 1-forms (see Wald \cite{Wald84}). 
Expressed in terms of $\Gamma_{\alpha \beta \gamma}$, the Lorentz gauge condition is
\begin{equation}
D_{\delta}\:\Gamma_{\alpha\beta}^{\:\:\:\:\:\delta}+\Gamma_{\alpha\beta}^{\:\:\:\:\:\gamma}\:\Gamma^{\:\delta}_{\:\:\:\gamma\delta}=0.
\end{equation}
The Lorentz gauge results in six evolution equations:
\begin{align}
\label{lorentzevoln}
D_{0}\:a_{b}-D_{c}\:K_{cb}=(a_{c}-2\:n_{c})\:K_{cb}-(Tr K)\:a_{b},\nonumber \\ 
D_{0}\:\omega_{b}+D_{c}\:N_{cb}=-(a_{c}-2\:n_{c})\:N_{cb}-(Tr K)\:\omega_{b},
\end{align}
and no additional constraint equations.  Note that the
principal terms in Eq.~(\ref{lorentzevoln}) are identical to those for the Nester gauge
evolution equations, Eq.~(\ref{nesterevoln}).

The evolution equations expressed in terms of directional derivatives, Eqs.~(\ref{kabevoln}), 
(\ref{nabevoln}), and (with a dynamic gauge) (\ref{nesterevoln}) or (\ref{lorentzevoln}),
can be written in a condensed notation,
\begin{equation}
\label{condensed}
D_{0}\:{\bf q} + \boldsymbol{M}^a\:D_a\:{\bf q}={\bf S},
\end{equation}
where ${\bf q}$ is a vector of the twenty-four variables $N_{ab},\:K_{ab},\:a_a,$ and $\omega_a$.  Furthermore, 
$\boldsymbol{M}^a$ are three sparse $24\times 24$ matrices whose only nonzero elements are $\pm\:1$, and ${\bf S}$ is a vector of source terms.  If one orders the variables so that
\begin{multline}
\label{q}
{\bf q}=(N_{11},N_{21},N_{31},a_1,K_{11},K_{21},K_{31},\omega_1,\\
\shoveleft{N_{12},N_{22},N_{32},a_2,K_{12},K_{22},K_{32},\omega_2,}\\
\shoveleft{N_{13},N_{23},N_{33},a_3,K_{13},K_{23},K_{33},\omega_3}),
\end{multline}
then the $\boldsymbol{M}^a$ matrices have a simple block diagonal structure.
Each is composed of three identical $8\times8$ blocks, respectively:
\begin{equation}
\boldsymbol{M}_{block}^1=
\begin{bmatrix}
	0&0&0&0&0&0&0&1 \\
	0&0&0&0&0&0&-1&0 \\
	0&0&0&0&0&1&0&0 \\
	0&0&0&0&-1&0&0&0 \\
	0&0&0&-1&0&0&0&0 \\
	0&0&1&0&0&0&0&0 \\
	0&-1&0&0&0&0&0&0 \\
	1&0&0&0&0&0&0&0 
\end{bmatrix},
\end{equation}
\begin{equation}
\boldsymbol{M}_{block}^2=
\begin{bmatrix}
	0&0&0&0&0&0&1&0 \\
	0&0&0&0&0&0&0&1 \\
	0&0&0&0&-1&0&0&0 \\
	0&0&0&0&0&-1&0&0 \\
	0&0&-1&0&0&0&0&0 \\
	0&0&0&-1&0&0&0&0 \\
	1&0&0&0&0&0&0&0 \\
	0&1&0&0&0&0&0&0 
\end{bmatrix},
\end{equation}
\begin{equation}
\label{M3}
\boldsymbol{M}_{block}^3=
\begin{bmatrix}
	0&0&0&0&0&-1&0&0 \\
	0&0&0&0&1&0&0&0 \\
	0&0&0&0&0&0&0&1 \\
	0&0&0&0&0&0&-1&0 \\
	0&1&0&0&0&0&0&0 \\
	-1&0&0&0&0&0&0&0 \\
	0&0&0&-1&0&0&0&0 \\
	0&0&1&0&0&0&0&0 
\end{bmatrix}.
\end{equation}

\section{\label{hyperbolic}Hyperbolic structure of the coordinate equations}

The beautiful quasi-FOSH tetrad
formulation with constant coefficients given in Eqs.~(\ref{condensed}) to (\ref{M3}) 
is not so simple, or so beautiful,
when expressed in terms of coordinates.  However, we proceed to show
that the coordinate equations are still quite
manageable and are, indeed, symmetrizable hyperbolic.

Substituting Eqs.~(\ref{D0}) and (\ref{Daagain}) into Eq.~(\ref{condensed}), we express the 
tetrad evolution equations in terms of partial 
derivatives along coordinate directions:
\begin{equation}
\label{timecoeffeqn}
\boldsymbol{T}\:D_0\:{\bf q}+\boldsymbol{M}^a\:B_a^k\:\frac{\partial \:{\bf q}}{\partial x^k}={\bf S},
\end{equation}
where 
\begin{equation}
\boldsymbol{T}\equiv\big[\boldsymbol{I}+\boldsymbol{M}^a\:{A}_a\big],
\end{equation}
and $\boldsymbol{I}$ is the identity matrix.  For clarity of notation, we let $D_0$ represent the partials
in Eq.~(\ref{D0}).  The $\boldsymbol{T}$ matrix has a block diagonal structure composed
of three identical $8 \times 8$ entries which are  
\begin{multline}
\label{timecoeffmatrix}
\boldsymbol{T}_{block}=\\
\begin{bmatrix}
	1&0&0&0&0&-{A}_3&{A}_2&{A}_1\\
	0&1&0&0&{A}_3&0&-{A}_1&{A}_2\\
	0&0&1&0&-{A}_2&{A}_1&0&{A}_3\\
	0&0&0&1&-{A}_1&-{A}_2&-{A}_3&0\\
	0&{A}_3&-{A}_2&-{A}_1&1&0&0&0\\
	-{A}_3&0&{A}_1&-{A}_2&0&1&0&0\\
	{A}_2&-{A}_1&0&-{A}_3&0&0&1&0\\
	{A}_1&{A}_2&{A}_3&0&0&0&0&1 
\end{bmatrix}.
\end{multline}
We now multiply Eq.~(\ref{timecoeffeqn}) by $\boldsymbol{T}^{-1}$ to give
\begin{equation}
\label{characteristic}
D_0\:{\bf q}+\boldsymbol{C}^a\:B_a^k\:\frac{\partial \:{\bf q}}{\partial x^k}=\boldsymbol{T}^{-1}{\bf S},
\end{equation}
where
\begin{equation}
\label{Ca}
\boldsymbol{C}^a\equiv\boldsymbol{T}^{-1}\boldsymbol{M}^a.
\end{equation}
$\boldsymbol{T}^{-1}$ is straightforward to calculate, since it is a block diagonal matrix with each block equal to the inverse of $\boldsymbol{T}_{block}$
in Eq.~(\ref{timecoeffmatrix}), where 
\begin{multline}
\boldsymbol{T}^{\:-1}_{block}=\frac{1}{1-A_a\:A_a}\:\times\\
\begin{bmatrix}
	1&0&0&0&0&{A}_3&-{A}_2&-{A}_1\\
	0&1&0&0&-{A}_3&0&{A}_1&-{A}_2\\
	0&0&1&0&{A}_2&-{A}_1&0&-{A}_3\\
	0&0&0&1&{A}_1&{A}_2&{A}_3&0\\
	0&-{A}_3&{A}_2&{A}_1&1&0&0&0\\
	{A}_3&0&-{A}_1&{A}_2&0&1&0&0\\
	-{A}_2&{A}_1&0&{A}_3&0&0&1&0\\
	-{A}_1&-{A}_2&-{A}_3&0&0&0&0&1 
\end{bmatrix}.
\end{multline}

The system given in Eq.~(\ref{characteristic}) is hyperbolic according to the
definition given in \cite{LeV02} if any linear combination $b_a\:\boldsymbol{C}^a$
of $\boldsymbol{C}^1$, $\boldsymbol{C}^2$, and $\boldsymbol{C}^3$ can be diagonalized with a complete set
of eigenvectors and real eigenvalues.  The linear combination for propagation in the $x^k$ coordinate
direction has $b_a=B_a^k$.  Solving the eigensystem of the combined $8\times8$ matrix in Mathematica
gives for the eigenvalues
\begin{equation}
\label{csumevals}
\frac{-\boldsymbol{b}\cdot\boldsymbol{A}\pm \sqrt{\boldsymbol{b}\cdot\boldsymbol{b}
-(\boldsymbol{b}\times\boldsymbol{A})\cdot(\boldsymbol{b}\times\boldsymbol{A})}}
{1-\boldsymbol{A}\cdot\boldsymbol{A}},
\end{equation}
where $\boldsymbol{b}$ and $\boldsymbol{A}$ are the 3-vectors in the spatial orthonormal frame
with components $b_a$ and $A_a$, respectively.  The dot and cross products are the standard
vector operations, and the upper/ lower signs on the square root are for left/ right-
propagating (relative to $\boldsymbol{b}$) modes.  The eigenvalues are real as long as
$\boldsymbol{A}\cdot\boldsymbol{A}=A_a\:A_a < 1$.  The eigenvectors (given in Appendix \ref{appB})
form a complete set.  The lapse and shift hidden in the $D_0$ operator 
modify the eigenvalues in a trivial way (multiply by the lapse $\alpha$, then subtract the
shift $\beta^k$), but have no effect on the eigenvectors.

For Eq.~(\ref{characteristic}) to also be symmetrizable by the definition of 
Lindblom and Scheel \cite{LS02}, a positive definite symmetric matrix must be found which 
multiplies the $\boldsymbol{C}^a$ matrices to give symmetric matrices.  The obvious candidate
for such a symmetrizer is $\boldsymbol{T}$, since $\boldsymbol{T}\:\boldsymbol{C}^a=\boldsymbol{M}^a$
by Eq.~(\ref{Ca}), and the $\boldsymbol{M}^a$ matrices are symmetric.  $\boldsymbol{T}$ is real and symmetric so a necessary and
sufficient condition for it to be positive definite is that all its eigenvalues
are positive \cite{Strang}.  This requires that 
$A_a\:A_a\:<\:1$.

Note that our saying the system is hyperbolic is contingent on the evolution of the $B_a^k$ components
being hyperbolic.  This is discussed further
in Sec. \ref{AaandBakeqns}.

\section{Evolution and constraint equations for ${A}_a$ and $B_a^k$}
\label{AaandBakeqns}

The commutator of the orthonormal basis vectors is
\begin{equation}
\label{commutator}
[\boldsymbol{e}_{\alpha},\:\boldsymbol{e}_{\beta}]=\nabla_{\alpha}\:\boldsymbol{e}_{\beta}-\nabla_{\beta}\:\boldsymbol{e}_{\alpha}=(\Gamma_{\;\beta\alpha}^{\gamma}-\Gamma_{\;\alpha\beta}^{\gamma})\:\boldsymbol{e}_{\gamma}.
\end{equation}
Expressed in terms of tetrad components and partial derivatives,
\begin{equation}
\label{commutcomps}
[\boldsymbol{e}_{\alpha},\:\boldsymbol{e}_{\beta}]=\Big(\lambda_{\alpha}^{\mu}\;\frac{\partial\lambda_{\beta}^{\nu}}{\partial x^{\mu}}-\lambda_{\beta}^{\mu}\;\frac{\partial\lambda_{\alpha}^{\nu}}{\partial x^{\mu}}\Big)\:\frac{\partial}{\partial x^{\nu}}.
\end{equation}
Expanding Eq.~(\ref{commutcomps}) and collecting terms gives
\begin{multline}
\label{lhs}
{\Big(\lambda_\alpha^t\;\frac{\partial\lambda_\beta^k}{\partial t}+\lambda_\alpha^k\;\frac{\partial\lambda_\beta^k}{\partial x^k}-\lambda_\beta^t\;\frac{\partial\lambda_\alpha^k}{\partial t}-\lambda_\beta^k\;\frac{\partial\lambda_\alpha^k}{\partial x^k}\Big)\:\frac{\partial}{\partial x^k}+} \\
\shoveleft{\Big(\lambda_\alpha^t\;\frac{\partial\lambda_\beta^t}{\partial t}+\lambda_\alpha^k\;\frac{\partial\lambda_\beta^t}{\partial x^k}-\lambda_\beta^t\;\frac{\partial\lambda_\alpha^t}{\partial t}-\lambda_\beta^k\;\frac{\partial\lambda_\alpha^t}{\partial x^k}\Big)\:\frac{\partial}{\partial t}}.
\end{multline}
Doing the same in Eq.~(\ref{commutator}), we get
\begin{multline}
\label{rhs}
\big[\big(\Gamma_{\;\beta\alpha}^0-\Gamma_{\;\alpha\beta}^0\big)\;\lambda_0^k+\big(\Gamma_{\;\beta\alpha}^c-\Gamma_{\;\alpha\beta}^c\big)\;\lambda_c^k\big]\;\frac{\partial}{\partial x^k}+\\
\shoveleft{\big[\big(\Gamma_{\;\beta\alpha}^0-\Gamma_{\;\alpha\beta}^0\big)\;\lambda_0^t+\big(\Gamma_{\;\beta\alpha}^c-\Gamma_{\;\alpha\beta}^c\big)\;\lambda_c^t\big]\;\frac{\partial}{\partial t}}.
\end{multline}
Set Eqs.~(\ref{lhs}) and (\ref{rhs}) equal, and let the index $\alpha=0$ and the index $\beta=a$.  
Simplifying, we obtain evolution equations for $B_a^k$ and ${A}_a$:
\begin{eqnarray}
\label{Bakevoln}
D_0\:B_a^k+\frac{B_a^l}{{\alpha}}\:\frac{\partial {\beta}^k}{\partial x^l} & = & \frac{1}{\alpha}\:\Big(\frac{\partial}{\partial t}-\mathcal{L}_\beta\Big)\:B_a^k \nonumber \\
 & = & -\varepsilon_{abc}\:\omega_b\:B_c^k-K_{ac}\:B_c^k,
\end{eqnarray}
\begin{equation}
\label{Aaevoln}
D_0\:{A}_a=a_a-\varepsilon_{abc}\:\omega_b\:{A}_c-K_{ac}\:{A}_c-\frac{B_a^l}{{\alpha}}\:\frac{\partial {\alpha}}{\partial x^l},
\end{equation}
where $\mathcal{L}_\beta$ in Eq.~(\ref{Bakevoln}) is the Lie derivative.
For fixed lapse and shift, the
evolution of both the $B_a^k$ and the $A_a$ is just advection along the
congruence worldlines and trivially hyperbolic.

The congruence can always be chosen to be orthogonal to
the initial hypersurface, so $A_a = 0$ initially.  However, $A_a$ will not
remain zero during the subsequent evolution unless the condition
\begin{equation}
\label{lapsecond}
B_a^l\:\frac{\partial}{\partial x^l}\:log\:\alpha=a_a
\end{equation}
is satisfied at all times.  For either the fixed or the
dynamic gauge conditions on $a_a$ considered here, the evolution of $B_a^l$
and/or $a_a$ is inconsistent with Eq.~(\ref{lapsecond}), except possibly for very
special initial conditions.

Repeating the same process as above, but with the index $\alpha=a$ and the index $\beta=b$, 
we obtain constraint equations for $B_a^k$ and ${A}_a$:
\begin{multline}
\label{Bakconstraint}
\varepsilon_{cab}\:B_a^l\:\frac{\partial B_b^k}{\partial x^l}=N_{dc}\:B_d^k- \\
\shoveleft{(TrN)\:B_c^k+\varepsilon_{cab}\:{A}_a\:(\varepsilon_{bdf}\:\omega_d\:B_f^k+K_{bd}\:B_d^k)},
\end{multline}
\begin{multline}
\label{Aaconstraint}
\varepsilon_{cab}\:B_a^l\:\frac{\partial {A}_b}{\partial x^l}=2\:\Omega_c+N_{dc}\:{A}_d-\\
\shoveleft{(TrN)\:{A}_c+\varepsilon_{cab}\:{A}_a\:(-a_b+\varepsilon_{bdf}\:\omega_d\:{A}_f+K_{bd}\:{A}_d)}.
\end{multline}
Note that if $A_a = 0$, the constraint of Eq.~(\ref{Aaconstraint}) is satisfied
automatically, and Eq.~(\ref{Bakconstraint}) can be used to calculate all of the $N_{ab}$
from the $B_a^k$.

\section{The Initial Value Problem}
\label{IVP}

Initial conditions for the variables must be chosen so all
relevant constraints are satisfied on the initial hypersurface.  This is
most easily accomplished if the initial tetrad is oriented so the tetrad
congruence is orthogonal to the initial hypersurface.  Then, 
$\Omega_a = 0$ and $A_a = 0$ initially.  The initial $B_a^k$ are the components of
an orthonormal triad of vectors lying in the initial hypersurface,
related to the inverse of the spatial metric of the hypersurface by $h^{kl}
= B_a^k\:B_a^l$.  One way to construct consistent initial conditions is to
solve the initial value problem using standard $3+1$ methods for the
spatial metric and the extrinsic curvature.  Construct orthonormal triad
fields by a Gramm-Schmidt orthogonalization procedure, orienting
the $B_1$ vector along the $x^1$ coordinate direction, and $B_2$ in the
$x^1-x^2$ plane, for instance.  Find the $N_{ab}$ from the commutators of the
$B_a$ 3-vectors using  Eq.~(\ref{Bakconstraint}).  The $K_{ab}$ are simply the projections of the
coordinate components of the extrinsic curvature as found from the $3+1$
initial value problem along the orthonormal triad vectors.  This
procedure is guaranteed to give consistent initial conditions for the
tetrad vectors, as long as the spatial coordinates are basically
Cartesian, {\it ie.}, there are no spatial coordinate singularities.  The
initial acceleration and angular velocity of the tetrad are arbitrary,
except in the context of the Nester gauge.  In this case, the initial
angular velocity $\omega_a$ and the initial $a_a - 2\:n_a$ must have vanishing
exterior derivatives. However, there is no guarantee that there are not
large twists of the initial triad vectors, possibly leading to large
gauge transients in the context of one of the dynamical gauge
conditions.

A more elegant and, likely, a better-behaved choice for the initial
spatial triad is to require that it satisfy the 3D Nester
gauge conditions in the hypersurface.  These conditions are that the 3D
one-form $\tilde{q}_{\:a}$, whose triad components are $\Gamma_{a\:\:\:\:b}^{\:\:\:b}$, has zero
exterior derivative, and that the trace of the $N_{ab}$ matrix vanish, 
{\it ie.} $N_{11}+ N_{22} + N_{33} = 0$.  
For simple topologies, the first condition is
equivalent to the condition that $2\:n_a = \varepsilon_{abc}\:N_{bc}$ be the gradient of a
scalar.  Finding a solution for the $B_a^k$ which satisfies these
conditions is, in general, a non-trivial elliptic problem (see \cite{JN89} and \cite{DMH89}).  
However, the situation is much
simpler for conformally flat 3-geometries.  Taking Cartesian basis
vectors in the conformal geometry, the conformal $N_{ab}$ are zero.  If
the conformal factor which generates the physical metric is $e^{4 \psi}$, the
conformal transformation simply rescales the $B_a^k$ by a factor $e^{-2\psi}$. 
The physical $n_a$ equal $-2\:D_a\:\psi$ (the gradient of a scalar), and the
symmetric part of $N_{ab}$ is still zero.  The 3D Nester condition is
satisfied.

\section{\label{conclusion}Discussion}

Although the emphasis of the tetrad formalisms in the literature is on evolving
spacetimes with physically defined flows, we think a tetrad formalism
may be useful for vacuum numerical calculations of black hole spacetimes.  The tetrad formulation
we have presented, based on that of Estabrook, Robinson, and Wahlquist
\cite{ERW97}, allows control of the evolution of the timelike congruence through either dynamic
or fixed gauge conditions on $a_a$ and $\omega_a$.  The lapse and shift we have
introduced allow for a completely arbitrary evolution of the coordinates
with minimal complication of the formalism. 
Furthermore, since the variables evolved are defined relative
to the orthonormal frames, the metric is the Minkowski metric and there are no
nonlinearities in the equations associated with the inverse metric. 
The system of equations based on coordinate derivatives is symmetrizable hyperbolic,
though admittedly more complicated than the quasi-FOSH system based
on directional derivatives.  Finally, the variables $N_{ab},\:K_{ab},\:a_a$ and $\omega_a$ are all scalars,
so derivatives of the shift only appear in the evolution equations for $B_a^k$.
We have successfully implemented this tetrad formalism 
for 1D colliding gravitational plane waves \cite{BuBa02},
where it results in substantially better accuracy and stability compared
with our calculations in the $3+1$ framework \cite{BaBu02}.

There are still important issues to be worked out in order
to apply this tetrad formalism to 3D black hole codes.
First, one must deal with the complications that arise when evolving
the congruence as well as the hypersurface. 
The congruence stays timelike as long as the acceleration is
bounded; however, in order for the the hypersurfaces to remain spacelike,
the condition $A_a\:A_a\:<\:1$ must be satisfied
at all times.  If $A_a\:A_a\:=\:1$, the system breaks down completely.
The hypersurfaces become null, causing the $3+1$ lapse to blow up.  In addition,
the coordinate equations
become singular (see Appendix \ref{appB}). 
$A_a\:A_a$ can be kept small by an adjustment of the lapse as the calculation
proceeds.  Such a resetting of the lapse invalidates theorems which bound the growth of the solution based on the symmetrizable 
hyperbolic structure of the system, but does not affect other advantages of the hyperbolic formulation,
such as dealing with boundary conditions.  

Second, attaining a stationary solution at late times for general 3D black hole numerical
calculations may be problematic in a tetrad formulation.  Such solutions are
potentially desirable for long-term stability.  With bounded acceleration, congruence
worldlines just outside the apparent horizon will necessarily cross the
horizon and be trapped by the black hole.  Worldlines further out may
actually accelerate outward relative to a static or stationary observer.
While a shift adjusted just right may keep grid stretching under control in
spite of this, the adjustment will have to be time-dependent except for very
special initial conditions on the congruence.  Knowing what these special
initial conditions are requires knowing the whole future solution ahead of
time, and thus is not practical.  

Both of these issues will probably best be addressed
by extending the symmetrizable hyperbolic system to allow for a dynamic lapse, 
to keep $A_a\:A_a$ small, and a dynamic shift, to control grid-stretching.

\appendix
\section{Source terms of tetrad evolution and quasi-constraint equations}
\label{appA}
The tetrad energy quasi-constraint equation, derived from $G_{00}=0$, is
\begin{multline}
2\:D_a\:n_a=-2\:\omega_a\:\Omega_a+N_{cd}\:N_{cd}+\\
\shoveleft{\frac{1}{2}(K_{cd}\:K_{dc}-(TrK)^2-N_{cd}\:N_{dc}-(TrN)^2)}.
\end{multline}
The tetrad momentum quasi-constraint equations, obtained from $G_{0a}=0$, are 
\begin{multline}
D_1\:(K_{22}+K_{33})-D_2\:K_{12}-D_3\:K_{13}\\
\shoveleft{=2\:\varepsilon_{1bc}\:a_b\:\Omega_c+\varepsilon_{1bc}\:K_{bd}\:N_{dc}-2\:K_{1c}\:n_c,}
\end{multline}
\begin{multline}
D_2\:(K_{11}+K_{33})-D_1\:K_{21}-D_3\:K_{23}\\
\shoveleft{=2\:\varepsilon_{2bc}\:a_b\:\Omega_c+\varepsilon_{2bc}\:K_{bd}\:N_{dc}-2\:K_{2c}\:n_c,}
\end{multline}
\begin{multline}
D_3\:(K_{11}+K_{22})-D_1\:K_{31}-D_2\:K_{32}\\
\shoveleft{=2\:\varepsilon_{3bc}\:a_b\:\Omega_c+\varepsilon_{3bc}\:K_{bd}\:N_{dc}-2\:K_{3c}\:n_c.}
\end{multline}
Analogous quasi-constraint equations for $\Omega_a$ and $N_{ab}$ are
\begin{multline}
2\:{D_a}\:\Omega_a=2\:a_a\:\Omega_a+4\:n_a\:\Omega_a,
\end{multline}
\begin{multline}
D_1\:(N_{22}+N_{33})-D_2\:N_{12}-D_3\:N_{13}=-2\:\varepsilon_{1bc}\:\omega_b\:\Omega_c+\\
\shoveleft{2\:K_{1c}\:\Omega_c-\varepsilon_{1bc}\:N_{1b}\:N_{c1}+2\:n_1\:(N_{22}+N_{33}),}
\end{multline}
\begin{multline}
D_2\:(N_{11}+N_{33})-D_1\:N_{21}-D_3\:N_{23}=-2\:\varepsilon_{2bc}\:\omega_b\:\Omega_c+\\
\shoveleft{2\:K_{2c}\:\Omega_c-\varepsilon_{2bc}\:N_{2b}\:N_{c2}+2\:n_2\:(N_{11}+N_{33}),}
\end{multline}
\begin{multline}
D_3\:(N_{11}+N_{22})-D_1\:N_{31}-D_2\:N_{32}=-2\:\varepsilon_{3bc}\:\omega_b\:\Omega_c+\\
\shoveleft{2\:K_{3c}\:\Omega_c-\varepsilon_{3bc}\:N_{3b}\:N_{c3}+2\:n_3\:(N_{11}+N_{22}).}
\end{multline}

To obtain the true constraint equations from these quasi-constraint equations, 
Eq.~(\ref{Da}) must be substituted for each
appearance of $D_a$.  Then Eq.~(\ref{characteristic})
must be used to eliminate the terms involving $D_0$ that have been introduced.
The true constraints are only needed to check accuracy as the solution evolves,
once consistent initial data is obtained.

The sources for $K_{ab}$ and $N_{ab}$ evolution Eqs.~(\ref{kabevoln}) and (\ref{nabevoln}) are
\begin{multline}
\label{SKab}
S\line(1,0){1}K_{ab}=a_{a}\:a_{b}+\varepsilon_{bcd}\:(-N_{ac}\:a_{d}+K_{ac}\:\omega_{d})+\\
\shoveleft{\varepsilon_{acd}\:K_{cb}\:\omega_{d}+\frac{1}{2}\:\varepsilon_{adf}\:\varepsilon_{bce}\:(K_{dc}\:K_{fe}-N_{dc}\:N_{fe})+}\\
\shoveleft{(Tr N)\:N_{ab}-N_{ca}\:N_{cb}-K_{ac}\:K_{cb}+2\:\omega_{b}\:\Omega_{a},}
\end{multline}
\begin{multline}
\label{SNab}
S\line(1,0){1}N_{ab}=-a_{a}\:\omega_{b}+\varepsilon_{bcd}\:(K_{ac}\:a_{d}+N_{ac}\:\omega_{d})+\\
\shoveleft{\varepsilon_{acd}\:N_{cb}\:\omega_{d}+\varepsilon_{adf}\:\varepsilon_{bce}\:N_{dc}\:K_{fe}-}\\
\shoveleft{(Tr N)\:K_{ab}+ N_{ca}\:K_{cb}-K_{ac}\:N_{cb}+2\:a_{b}\:\Omega_{a}.}
\end{multline}

\section{Details of coordinate equations}
\label{appB}

In this appendix, we expand Eq.~(\ref{characteristic}).  Because of the block diagonal
structure of the $\boldsymbol{C}^a$ and $\boldsymbol{M}^a$ matrices, it is only necessary to work with one set of
eight variables.  We will let the free index $d= 1,\:2,\:$ or $3$ 
(for the first, second, and third set of eight variables in the vector {\bf q}), {\it ie.}
\begin{equation}
{\bf q}_d=(N_{1d},\:N_{2d},\:N_{3d},\:a_d,\:K_{1d},\:K_{2d},\:K_{3d},\:\omega_d).
\end{equation}
So as to include a non-zero shift with minimal notation, we use $D_0$ to represent the 
partial derivatives in Eq.~(\ref{D0}).  Then we can write Eq.~(\ref{characteristic}) as
\begin{multline}
D_0\:{\bf q}_d+\\
\shoveleft{[\boldsymbol{C}^1_{block}\:B_1^k+\boldsymbol{C}^2_{block}\:B_2^k+\boldsymbol{C}^3_{block}\:B_3^k]\:\frac{\partial {\bf q}_d}{\partial x^k}={\bf S}_d,}
\end{multline}
where
\begin{multline}
\boldsymbol{C}_{block}^1=-\frac{1}{1-A_a\:A_a}\:\times\\
\begin{bmatrix}
	{A}_1&-{A}_2&-{A}_3&0&0&0&0&-1\\
	{A}_2&{A}_1&0&-{A}_3&0&0&1&0\\
	{A}_3&0&{A}_1&{A}_2&0&-1&0&0\\
	0&{A}_3&-{A}_2&{A}_1&1&0&0&0\\
	0&0&0&1&{A}_1&-{A}_2&-{A}_3&0\\
	0&0&-1&0&{A}_2&{A}_1&0&-{A}_3\\
	0&1&0&0&{A}_3&0&{A}_1&{A}_2\\
	-1&0&0&0&0&{A}_3&-{A}_2&{A}_1
\end{bmatrix},
\end{multline}
\begin{multline}
\boldsymbol{C}_{block}^2=-\frac{1}{1-A_a\:A_a}\:\times\\
\begin{bmatrix}
	{A}_2&{A}_1&0&{A}_3&0&0&-1&0\\
	-{A}_1&{A}_2&-{A}_3&0&0&0&0&-1\\
	0&{A}_3&{A}_2&-{A}_1&1&0&0&0\\
	-{A}_3&0&{A}_1&{A}_2&0&1&0&0\\
	0&0&1&0&{A}_2&{A}_1&0&{A}_3\\
 	0&0&0&1&-{A}_1&{A}_2&-{A}_3&0\\
	-1&0&0&0&0&{A}_3&{A}_2&-{A}_1\\
	0&-1&0&0&-{A}_3&0&{A}_1&{A}_2
\end{bmatrix},
\end{multline}
\begin{multline}
\boldsymbol{C}_{block}^3=-\frac{1}{1-A_a\:A_a}\:\times\\
\begin{bmatrix}
	{A}_3&0&{A}_1&-{A}_2&0&1&0&0\\
	0&{A}_3&{A}_2&{A}_1&-1&0&0&0\\
	-{A}_1&-{A}_2&{A}_3&0&0&0&0&-1\\
	{A}_2&-{A}_1&0&{A}_3&0&0&1&0\\
	0&-1&0&0&{A}_3&0&{A}_1&-{A}_2\\
	1&0&0&0&0&{A}_3&{A}_2&{A}_1\\
 	0&0&0&1&-{A}_1&-{A}_2&{A}_3&0\\
	0&0&-1&0&{A}_2&-{A}_1&0&{A}_3
\end{bmatrix},
\end{multline}
and
\begin{multline}
\label{Sd}
{\bf S}_d=\frac{1}{1-A_a\:A_a}\:\times\\
\begin{bmatrix}
{A}_3\:S\line(1,0){1}K_{2d}-{A}_2\:S\line(1,0){1}K_{3d}+S\line(1,0){1}N_{1d}-{A}_1\:S\line(1,0){1}\omega_d\\
-{A}_3\:S\line(1,0){1}K_{1d}+{A}_1\:S\line(1,0){1}K_{3d}+S\line(1,0){1}N_{2d}-{A}_2\:S\line(1,0){1}\omega_d\\
{A}_2\:S\line(1,0){1}K_{1d}-{A}_1\:S\line(1,0){1}K_{2d}+S\line(1,0){1}N_{3d}-{A}_3\:S\line(1,0){1}\omega_d\\
S\line(1,0){1}a_d+{A}_1\:S\line(1,0){1}K_{1d}+{A}_2\:S\line(1,0){1}K_{2d}+{A}_3\:S\line(1,0){1}K_{3d}\\
{A}_1\:S\line(1,0){1}a_d+S\line(1,0){1}K_{1d}-{A}_3\:S\line(1,0){1}N_{2d}+{A}_2\:S\line(1,0){1}N_{3d}\\
{A}_2\:S\line(1,0){1}a_d+S\line(1,0){1}K_{2d}+{A}_3\:S\line(1,0){1}N_{1d}-{A}_1\:S\line(1,0){1}N_{3d}\\
{A}_3\:S\line(1,0){1}a_d+S\line(1,0){1}K_{3d}-{A}_2\:S\line(1,0){1}N_{1d}+{A}_1\:S\line(1,0){1}N_{2d}\\
-{A}_1\:S\line(1,0){1}N_{1d}-{A}_2\:S\line(1,0){1}N_{2d}-{A}_3\:S\line(1,0){1}N_{3d}+S\line(1,0){1}\omega_d
\end{bmatrix}.
\end{multline}
The expressions for $S\line(1,0){1}N_{1d},\:S\line(1,0){1}N_{2d},\:S\line(1,0){1}N_{3d},\:S\line(1,0){1}K_{1d}$, $S\line(1,0){1}K_{2d},\:S\line(1,0){1}K_{3d}$ in Eq.~(\ref{Sd}) are obtained from Eqs.~(\ref{SKab}) and (\ref{SNab}).  Those for $\:S\line(1,0){1}a_d$ and $S\line(1,0){1}\omega_d$
are from either Eqs.~(\ref{nesterevoln}) or Eqs.~(\ref{lorentzevoln}).

The eight eigenvectors of the arbitrary linear combination $b_a\:\boldsymbol{C}^a_{block}$
consist of four pairs of left and right-propagating modes.  Each pair only involves one of
the four variables $K_{1d}$, $K_{2d}$, $K_{3d}$, $\omega_d$.  For propagation in the $x^k$
coordinate direction, $b_a=B_a^k$.  As in Sec. \ref{hyperbolic}, we simplify notation by using
$\boldsymbol{b}$ and $\boldsymbol{A}$ to denote the 3-vectors in the spatial orthonormal
frame with components $b_a$ and $A_a$.  Then the $\omega_d$ eigenvectors, normalized so
$\omega_d=\boldsymbol{b}\cdot\boldsymbol{b}={\left | \boldsymbol{b}\right |}^2$,
have $N_{bd}=[\boldsymbol{b}\times(\boldsymbol{b}\times\boldsymbol{A})]_b \pm b_b\:
\sqrt{{\left | \boldsymbol{b}\right |}^2-{\left | \boldsymbol{b}\times \boldsymbol{A}\right |}^2}$,
and $a_d=0$. The $K_{ad}$ eigenvectors, normalized so
$K_{ad}={\left | \boldsymbol{b}\right |}^2$,
have $N_{bd}=\varepsilon_{abc}\left([\boldsymbol{b}\times(\boldsymbol{b}\times\boldsymbol{A})]_c \pm b_c\:
\sqrt{{\left | \boldsymbol{b}\right |}^2-{\left | \boldsymbol{b}\times \boldsymbol{A}\right |}^2}\right)$
and $a_d=-\left([\boldsymbol{b}\times(\boldsymbol{b}\times\boldsymbol{A})]_a \pm b_a\:
\sqrt{{\left | \boldsymbol{b}\right |}^2-{\left | \boldsymbol{b}\times \boldsymbol{A}\right |}^2}\right)$. 
The upper sign on the square root corresponds to the upper sign in Eq.~(\ref{csumevals}) for the
eigenvalues.  Note that $N_{ad}$ is zero in the eigenvectors.

\begin{acknowledgments}
LTB and JMB gratefully acknowledge F. B. Estabrook and H. D. Wahlquist for their 
contributions to this work.  We would also like to 
thank L. Lindblom, M. Scheel, and D. Meier for their support and insightful discussions during
the CalTech visitors program.  
LTB was supported by the NASA Graduate Student Researchers Program
under Grant No. NGT5-50298 for the duration of this research.
\end{acknowledgments}

\bibliography{Tetrad_paper1}

\end{document}